\documentclass[a4paper,
               ]{jacow}
\usepackage{pdfpages,multirow,ragged2e} %
\makeatletter%
	\ifboolexpr{bool{xetex}}
	 {\renewcommand{\Gin@extensions}{.pdf,%
	                    .png,.jpg,.bmp,.pict,.tif,.psd,.mac,.sga,.tga,.gif,%
	                    .eps,.ps,%
	                    }}{}
\makeatother

\ifboolexpr{bool{xetex} or bool{luatex}} % test for XeTeX/LuaTeX
 {}                                      % input encoding is utf8 by default
 {\usepackage[utf8]{inputenc}}           % switch to utf8

\usepackage[USenglish]{babel}

\ifboolexpr{bool{jacowbiblatex}}%
 {%
  \addbibresource{jacow-test.bib}
  \addbibresource{biblatex-examples.bib}
 }{}
\begin{document}

\title{Ionization profile monitors for the IOTA proton beam \thanks{This manuscript has been authored by Fermi Research Alliance, LLC under Contract No. DE-AC02-07CH11359 with the U.S. Department of Energy, Office of Science, Office of High Energy Physics.}}

\author{H. Piekarz \thanks{hpiekarz@fnal.gov}, A. Romanov, R. Thurman-Keup, Fermilab, Batavia, IL 60510, USA \\
		V. Shiltsev, Northern Illinois University, DeKalb, IL 60115, USA} 
%		P. Contributor\textsuperscript{1}, Name of Institute or Affiliation, City, Country \\
%		\textsuperscript{1}also at Name of Secondary Institute or Affiliation, City, Country}
	
\maketitle

\begin{abstract}
Ionization profile monitors (IPMs) are widely used in accelerators for non-destructive and fast diagnostics of high energy particle beams. Two such monitors - one vertical and one horizontal - are being developed for installation in the IOTA storage ring at Fermilab. They will be used for turn-by-turn (microseconds scale) measurements of the 70 MeV/c IOTA proton beam sizes.
In this paper we present the IPMs design (largely following the FNAL Booster IPMs which employ no external guiding magnetic fields), their mechanical, vacuum, and electric subsystems and DAQ, and discuss anticipated effects on the beams circulating in IOTA. 
\end{abstract}

\section{Introduction}

%\label{intro}
Particle accelerators heavily rely on precise diagnostics and control of critical beam parameters such as intensity, pulse structure, position, transverse and longitudinal beam sizes, halo, etc  \cite{Strehl}. IOTA (Integrable Optics Test Accelerator) at Fermilab is a 40-m circumference ring dedicated to accelerator R\&D \cite{IOTA}. It is capable of operation with 100-150 MeV/c electrons and 70 MeV/c protons and requires flexible and precise beam diagnostics for a variety of beam experiments. Ionization profile monitors (IPMs) are considered as a primary tool for fast and accurate measurements of the IOTA proton beam profiles.  

\begin{figure}[!htb]
   \centering
   \includegraphics*[width=.8\columnwidth]{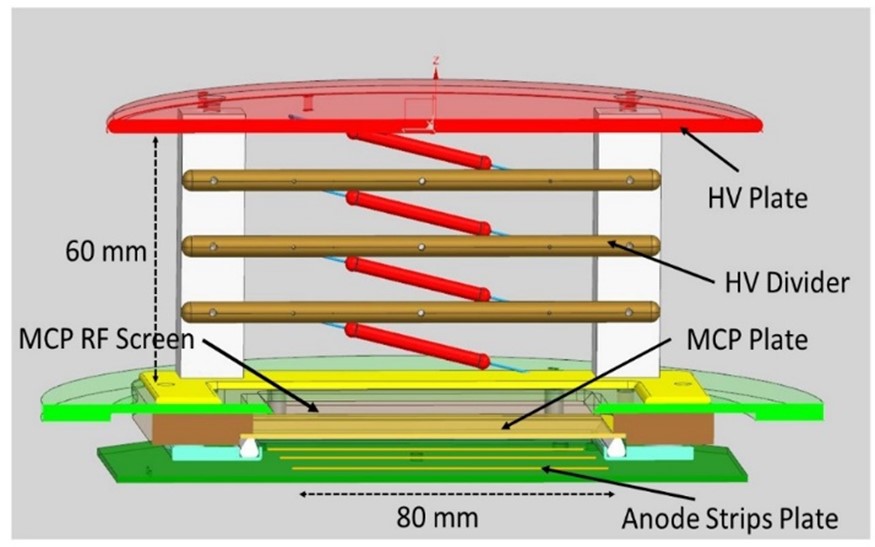}
   \caption{Close-up view of the IOTA IPMs. Anode strips at the bottom are parallel to the proton beam (not shown). }
   \label{fig:fig1}
\end{figure}

IPMs  \cite{Hornstra, Weisberg, Hochadel, Anne, Connolly, Jansson} are fast and non-destructive diagnostic tools used in proton and ion linacs, colliders and rapid cycling synchrotrons (RCS) \cite{Benedetti, Moore,  Levasseur, Eldred}. 
They operate by collecting ions or electrons created after the ionization of residual vacuum molecules by high energy charged particle beams \cite{Strehl, Wittenburg}, which are then guided to a detector by a uniform external electric field $E_{\rm ext}$. The detector is usually made of many thin parallel strips, whose individual signals are registered to make the beam profile signal ready for processing  -- see Fig.\ref{fig:fig1}. 

Space-charge forces of the primary beams make the measured IPM profiles different from those of the beams and must be correctly accounted. Theory of the IPM without a guiding magnetic field \cite{ShiltsevIPM} allows reconstruction of the actual beam sizes from measured IPM profiles given the key parameters, such as high-energy beam intensity $N$ and IPM extracting field $E_{\rm ext}=V_0/D$ which is the electric field due to the voltage gradient $V_0$ across the IPM gap $D$. It was shown that the space-charge expansion in IPM results in {\it proportional magnification of the profile} of the distribution of the secondary particles, and the rms transverse size of the IPM profile at the time when the secondary particle reaches the IPM detector is :   
\begin{equation}
    \sigma_m=\sigma_0 \cdot h \approx \sigma_0  \cdot \Big[ 1 + 
    \frac{2 U_{SC}}{E_{\rm ext} \sigma_0} \Big(\frac{ \Gamma({1 \over 4})}{3} \sqrt{d \over \sigma_0}  - {\sqrt{\pi} \over 2} \Big) 
\Big] \, .
\label{EQ1}
\end{equation} 
Here, the gamma-function $\Gamma({1 \over 4}) \approx 3.625$. The space-charge expansion factor $h$ is determined only by the space-charge potential of the primary proton beam $U_{SC}=30[V]J_b/\beta_p$, its rms size $\sigma_0$, the IPM extracting field $E_{\rm ext}=V_0/D$, and the beam-to-MCP distance $d$ but {\it it does not depend on the type of secondary species} (their mass and charge, etc).  Equation~\eqref{EQ1} can be easily solved, and the original $\sigma_0$ can be found from $\sigma_m$ \cite{ShiltsevIPM}.  

For the IOTA proton ring parameters - see Table \ref{tab:params} - Eq.\ref{EQ1} predicts very modest space-charge expansion of the beam profile in the IPMs $h-1 \approx 0.01-0.1$ and, therefore, the monitors can operate only with guiding electric field. No need for the guiding magnetic fields significantly simplifies the IPM design. 

\begin{figure}[!htb]
   \centering
   \includegraphics*[width=.99\columnwidth]{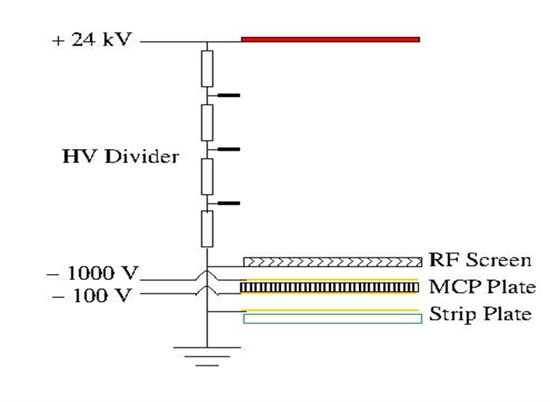}
   \caption{Conceptual view of IOTA IPM electric scheme.}
   \label{fig:fig2}
\end{figure}

\section {IPM SUBSYSTEMS}

\subsection{HV cage}
In the IOTA ring, the several mm wide proton beam with an average current of up to 8 mA will pass through the IPM HV cage - see Fig.\ref{fig:fig1}. The cage dimensions are 60 mm(height)$\times$60 mm(width)$\times$100 mm(length). The voltage on the upper plate is +16 kV (can vary upto 30 kV), the electric field uniformity is arranged by a four-stage voltage divider attached to copper bars (in brown). 
The positive ions of the residual gas ionized in the IPMs high electric field are pulled away from the positively charged electrode toward the grounded RF screen - see at the bottom of Figs. \ref{fig:fig1} and \ref{fig:fig2}, in gold. The negative potential (e.g., 1 kV) of the micro-channel plate (MCP) pulls the ions towards the MCP plate where they generate electrons. The latter, after amplification in the MCP, exit and proceed for another 7.5 mm to an array of thin, parallel anode strips spaced 0.5 mm apart at +100 V above the exit of the MCP, where the electrons are collected for further processing. The anode detector readout strip plate dimensions are 40 mm(width)$\times$80 mm(length) with 60-80 strips in each H and V monitor.

\subsection{Electronics and DAQ}
The 1.8 $\mu$s revolution period of IOTA is close to that of the Booster (2.2 $\mu$s) and the bandwidth of the existing Booster IPM electronics should be sufficient for turn-by-turn profile measurements in IOTA. The IOTA IPMs readout system will be based on that for the FNAL Booster IPMs \cite{Randy}. Both horizontal and vertical IPMs will each need: a) NIM crate with pre-amps and low-pass filters; b) VME readout system, similar to the fast ($\mu$s) digitizers built for the Booster IPMs; c) LabView control system (similar to the one for Booster IPMs), connected to the IOTA ACNET (later, EPICS) control system.    

\begin{figure}[!htb]
   \centering
   \includegraphics*[width=.99\columnwidth]{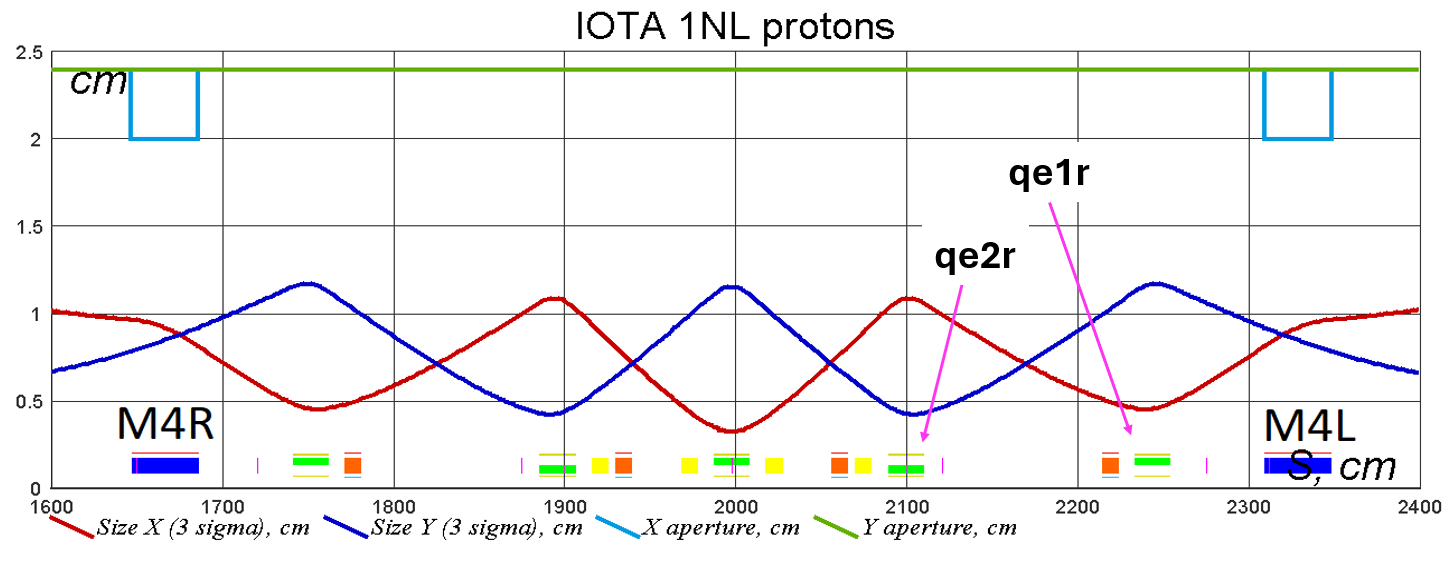}
   \caption{Projected IOTA proton beam size 3$\sigma_{x,y}$ at the IPMs location $S=2120-2210$ cm (red- horizontal, blue - vertical).}
   \label{fig:fig3}
\end{figure}

\subsection{Location and geometry}
\begin{figure*}[!htb]
   \centering
   \includegraphics*[width=.85\textwidth]{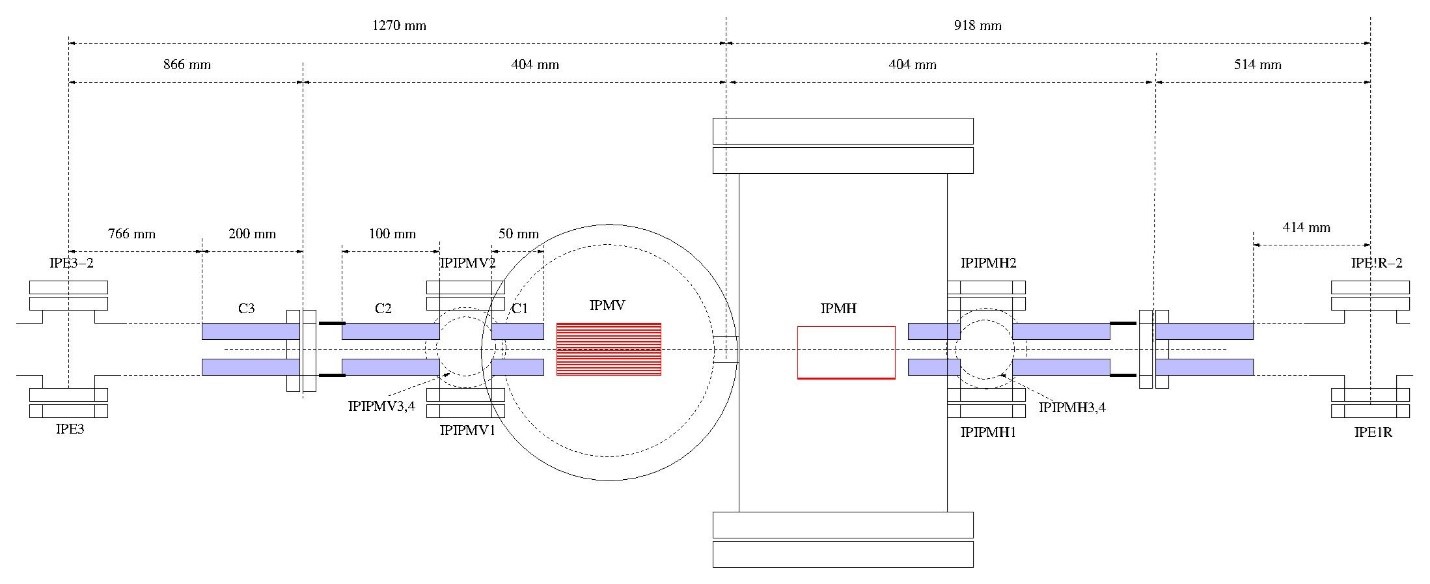}
   \caption{Schematic of vacuum system arrangement for IOTA IPMs insert.}
   \label{fig:fig4}
\end{figure*}

Both IPMs will be set side-by-side at the IOTA ring straight $\sim$0.8 m long section between quadrupole magnets $qe1r$ and $qe2r$ where the proton beam size is projected to be about 6-8 mm (three times the rms size - see Fig. \ref{fig:fig3}). Fig. \ref{fig:fig4} presents a schematic layout of the of the IOTA IPM insert.

\begin{table}[!hbt]
   \centering
   \caption{Main Parameters of IOTA IPMs }
   \begin{tabular}{lccc}
       \toprule
       \textbf{Parameter} & \textbf{Units}                      & \textbf{Design} & \textbf{Range} \\
       \midrule
{\it $\quad \quad$ IOTA beam} & & & \\
Proton {\it pc} & MeV & 68.5 & \\
Number of bunches & & 4 & 1-4 \\
Rev. period $T_r$ & $\mu$s & 1.83 & \\
RF frequency $f_{RF}$ & MHz & 2.19 & \\
Avg. current $I_b$ & mA & 2 & 0.5-8 \\
Tr. emitt. geom. $\varepsilon_{x,y}$& $\mu$m rms & 3 & 1-4 \\
Rms size at IPM $\sigma_{x,y}$ & mm & 2.5 & 1-5 \\
SC potential $U_{SC}$ & V & 0.9 & 0.2-3.5 \\
Mom.spread $\Delta p/p$ & rms & 2$\cdot 10^{-3}$ & (1-2)$\cdot 10^{-3}$ \\
Rms bunch length $\sigma_l$ & m & 1.7 & 1-2 \\
SC tuneshift $- \Delta Q_{SC}$ & 0.5 & 0.05-1 \\
Avg. vacuum $P_{IOTA}$ & Torr & 6$\cdot 10^{-10}$ &  $10^{-10...-8}$ \\
Nucl. lifetime $\tau_{vac}$ & s & 300 & 20-10$^3$ \\
SC lifetime $\tau_{SC}$ & s & <1 & 10$^{-3}$-2 \\
\midrule
{\it $\quad \quad$ IOTA IPMs} & & & \\
Number of IPMs &  & 2 & \\
Total length & m & 0.808 & \\
Min aperture & mm & 30 & \\
HV gap $D$ & mm & 60 & \\
HV cage length & mm & 100 & \\
Strips/IPM & & 60 & \\
Strip alignment & mrad & $\pm$1 & \\
Pitch $\Delta$ & mm & 0.5 & \\
IPM Voltage $V_0$ & kV & 16 & 4-30 \\
Vacuum $P_{IPM}$ & Torr & 6$\cdot 10^{-9}$ &  $10^{-9...-7}$ \\
Eff.vac.length & m & 0.48 & \\
Integr. time & turns & 1 & 1-10 \\
Ions/turn & 10$^3$ & 1.8 & 0.3-30 \\
SC expansion $h-1$ & & 0.03 & 0.01-0.1 \\
%           Right       & \SI{20}{mm} (\SI{0.79}{in})    %        & \SI{1.02}{in} (\SI{26}{mm})        \\
       \bottomrule
   \end{tabular}
   \label{tab:params}
\end{table}

\subsection{Vacuum system}
The IOTA vacuum beam pipe (1-7/8" or 47.6 mm inner diameter) maintains up to $\sim$10$^{-10}$ Torr vacuum pressure (that's the minimum expected average value, with the design specification of $6 \cdot 10^{-10}$ Torr). With such a pressure and the circumference of 40 m, the IOTA ring integrated vacuum load is $40 \cdot 10^{-10}$ Torr-m. The IPMs insert adds less than 10\% to that value. This means that for the 2 m long IPMs insert the injected gas pressure should not exceed $2 \cdot 10^{-10}$ Torr. The total length of the IPMs and associated vacuum equipment is 2.2 m (between the nearest two ion pumps). As the proper IPMs chambers occupy 0.404 m of space they use all allowable integral vacuum load with $\sim 10^{-9}$ Torr pressure of the injected gas. This situation requires that the pressure rise in the IPMs insert sections outside the IPMs vacuum chamber is strongly suppressed. For that, not only strong vacuum pumping outside the IPMs chambers is needed but reduction of the IPMs gas flow toward IPE3 and IPE1R ion pumps as well. Such a reduction is arranged by installation of aperture-restricting inserts ("collimators") inside the IPM vacuum system as shown in Fig.\ref{fig:fig4}. 

There are 4 new ion pumps to be installed right next to the left and to the right of the H and V IPMs, and 2 existing ion pumps at the IPE3 and IP1R locations will be replaced with much more powerful pumps to increase the pumping speeds of nitrogen and argon by a factor of 2.6 and 7, respectively. The flow from the IPMs space into the IOTA beam pipe is constrained with the series of “collimators”, C1, C2, C3. The total length of these collimators is 350 mm. Based on the projected proton beam sizes, the collimator orifice of 30 mm (diameter) was chosen. The pressures and the integrated pressures for the selected locations relevant to the IPMs insert were estimated using conductance formulae for the molecular gas flow regime. We calculate that a $ 10^{-9}$ Torr pressure inside the IPM cage  - arranged by a controlled leak - will result in a quite acceptable integrated pressure load of less than $5 \cdot 10^{-10}$ Torr-m - a small fraction of the total IOTA ring integral. The IPM vacuum pressure will be controlled by the variable leak valve (VAT, Series 59.0). 

There are indications that use of noble gases such as Ar, Kr, or Xe can be beneficial for IPM operation as they allow a smaller profile smearing \cite{Egberts, Egberts2}. Our analysis of the IOTA IPMs vacuum systems shows no significant difference in the resulting pressures between, e.g., Nitrogen and Argon. This is because as the pumping speed for argon gets lower its conductance through the pipes also gets lower leading to a pressure rise at the pipe ends in a similar range to that for nitrogen. We therefore, conclude other gases, such as Ne and Xe, can be successfully explored though their pumping speeds are likely lower than that for Argon. 

%\begin{figure}[!htb]
%   \centering
%   \includegraphics*[width=.9\columnwidth]{WEPG039_FIG3.png}
%   \caption{Projected IOTA proton beam 3$\sigma$ sizes in cm at the IPM location (red - horizontal, blue - verstical).}
%   \label{fig:fig3}
%\end{figure}

At the $10^{-9}$ Torr pressure of the gas injected into the IPM HV cage, the estimated number of generated ions by the IOTA beam of $9 \cdot 10^{10}$ protons at 70 MeV/c is $\sim$1200 per 1.83 $\mu$s turn, that is expected to be sufficient to provide statistically significant turn-by-turn proton beam profile measurement. 

\flushcolsend

\section{Effects on IOTA beams}

{\it General:} IOTA IPMs should be compatible with other IOTA experiments and plans (including the plan to make IOTA optics periodic with period $P$=2).  The IPMs can not be fully transparent to the IOTA operations: they generate transverse kicks in both H and V planes, they contribute about 10\% to the total vacuum load, the IPM vacuum system reduces the transverse aperture to 30 mm diameter (15 mm radius), and, in the end, the system of two monitors will take about 0.8 m (flange-to-flange) of the ring circumference.

{\it Effects on the 70 MeV/c proton beam orbit} can be estimated as: with voltage of 15 kV (the nominal operating voltage) over the 60mm IPM cage gap and 100 mm length, will result in about 5.1 mrad of angular kick (in both planes) and unacceptably large IOTA beam and orbit distortions. Compensation of these deflections will require $Bl \approx$ 12 Gauss-meter magnetics (the IOTA proton ring rigidity is $B\rho$=0.23 Tm=2300 Gauss-meter). There are two opportunities for this. The IOTA orbit correction system could be used to control/correct such distortions. The ring has 20 combined H/V/skew-quad correctors, with maximum dipole strength of 25 Gauss-meter (built with 270-turn air-cooled coils: 2A/5V). The most impactful could be the nearby (to the IPM) local correctors $(x,y)$SQE1R and $(x,y)$SQE2R. The second option involves installing new magnetic correctors along approximately 100mm-long C2 sections of the beam pipe, adjacent to the vacuum pump four-way ports IPIPMV1-4 and IPIPMH1-4, both upstream and downstream of the IPMs
– see Fig.\ref{fig:fig4}. 
%
%{\it Effects on the vacuum and transverse aperture restrictions} have been discussed above.

\section{CONCLUSION AND OUTLOOK}
We have designed the ionization profile monitors (IPMs) for the IOTA ring turn-by-turn proton beam profile measurements, and have started construction of the monitors. The IPMS are expected to be fully compatible with the IOTA 70 MeV/c proton operation. The IPMs are anticipated to be ready for installation in the ring early 2025, to be followed by a reasonable time for commissioning and initial tests. Of note, testing the IPMs with a pencil-like narrow electron beam - easily available at IOTA - will allow precise determination of the instrument point spread function (PSF) and its dependence on beam current and size – that information will be very helpful for the monitors operation with protons.

%\section{ACKNOWLEDGEMENTS}
%Any acknowledgement should be in a separate section directly preceding
%the \textbf{REFERENCES} or \textbf{APPENDIX} section.
%
%
% only for "biblatex"
%
\ifboolexpr{bool{jacowbiblatex}}%
	{\printbibliography}%

\begin{thebibliography}{99} % Use for 1-9 references
%\bibitem{Shiltsev} V. Shiltsev, Phys.Today, 73 (2020).
%	\bibitem{Costmodel} % V.Shiltsev, JINST 9 (2014) T07002.
%	V. Shiltsev, ``A phenomenological cost model for high energy particle accelerators'', \emph{Journal of Instrumentation}, vol. 9, p. T07002, 2013. \url{doi:10.1088/1748-0221/9/07/T07002}

%1
\bibitem{Strehl} P.~Strehl, {\it Beam Instrumentation and Diagnostics, vol. 120}, Berlin, Germany: Springer, 2006.

%2
\bibitem{IOTA} S.~Antipov et al, “IOTA (Integrable Optics Test Accelerator): facility and experimental beam physics program“, \emph{JINST} 12 T03002, 2017.
\url{doi:10.1088/1748-0221/12/03/T03002}. 

%3
\bibitem{Hornstra} F.~Hornstra, W. H.~DeLuca, ``Nondestructive beam profile detection systems for the Zero Gradient Synchrotron", in \textit{Proc. VI Conference on High Energy Acceleration}, Cambridge, MA, 1967, pp. 374--377.

%4
\bibitem{Weisberg} H.~Weisberg \textit{et al.}, ``An Ionization Profile Monitor for the Brookhaven AGS", \emph{IEEE Trans.Nucl.Sci.}, vol. 30, p. 2179, 1983. \url{doi:10.1109/TNS.1983.4332753}

%5
\bibitem{Hochadel} B.~Hochadel \textit{et al.}, ``A residual-gas ionization beam profile monitor for the Heidelberg Test Storage Ring TSR", \emph{Nucl.Instr.Meth. A}, vol. 343, p. 401, 1994. \url{doi:10.1016/0168-9002(94)90217-8}

%6
\bibitem{Anne} R.~Anne \textit{et al.}, ``A Noninterceptive heavy ion beam profile monitor based on residual gas ionization", \emph{Nucl.Instr.Meth. A}, vol. 329, p. 21, 1993.
\url{doi:10.1016/0168-9002(93)90918-8}

%7
\bibitem{Connolly} R.~Connolly \textit{et al.}, ``Beam profile measurements and transverse phase-space reconstruction on the Relativistic Heavy-Ion Collider", \emph{Nucl.Instr.Meth. A}, vol. 443, p. 2015, 2000. \url{doi:10.1016/S0168-9002(99)01162-6}

%8
\bibitem{Jansson} A.~Jansson \textit{et al.}, ``The Tevatron Ionization Profile Monitors", \emph{AIP Conf. Proc.}, vol. 868, p. 159, 2006. \url{doi:10.1063/1.2401401}

%9
\bibitem{Benedetti} F.~Benedetti \textit{et al.}, ``Design and development of Ionization Profile Monitor for the Cryogenic sections of the ESS Linac", \emph{EPJ Web of Conf.}, vol. 225, p. 01009, 2020. \url{doi:10.1051/epjconf/202022501009}

%10
\bibitem{Moore} R. S.~Moore, A.~Jansson, and V.~Shiltsev, ``Beam instrumentation for the Tevatron collider", \emph{J. Instrumentation}, vol. 4, no. 12, p. 12018, 2009. \url{doi:10.1088/1748-0221/4/12/P12018}

%11
\bibitem{Levasseur} S.~Levasseur \textit{et al.}, ``Development of a rest gas ionisation profile monitor for the CERN Proton Synchrotron based on a Timepix3 pixel detector", \emph{J. Instrumentation}, vol. 12, no. 02, p. C02050, 2017.
\url{doi:10.1088/1748-0221/12/02/C02050}

%12
\bibitem{Eldred} J.Eldred \textit{et al.}, ``Beam intensity effects in Fermilab Booster synchrotron", \emph{Phys.Rev.Accel.Beams}, vol. 24, p. 044001, 2021.
\url{doi:10.1103/PhysRevAccelBeams.24.044001}

%13
\bibitem{Wittenburg} K.~Wittenburg, ``Specific instrumentation and diagnostics for high-intensity hadron beams", in \textit{CAS - CERN Accelerator School: High Power Hadron Machines}, Mar. 2013, pp. 251--308. \url{doi:10.5170/CERN-2013-001.251}

%14
\bibitem{ShiltsevIPM} V.~Shiltsev, ``Space-charge effects in ionization beam profile monitors", \emph{Nucl.Instr.Meth. A}, vol. 986, p. 164744, 2021.
\url{doi:10.1016/j.nima.2020.164744}

%15
\bibitem{Randy} R.~Thurman-Keup, Internal Note FNAL-BEAMS-DOC-9775-v4 (unpublished, 2023).

%16
\bibitem{Egberts} J. Egberts et al., ”Detailed Experimental Characterization of an Ionization Profile Monitor,  Proc. 10th European Workshop on Beam Diagnostics and Instrumentation for Particle Accelerators (DIPAC’11), Hamburg, Germany (May 2011).

\bibitem{Egberts2} J.~Egberts, ``IFMIF-LIPAc Beam Diagnostics. Profiling and Loss Monitoring Systems", Ph.D. thesis, Universite Paris Sud, Orsay, France, 2012.


%14
%\bibitem{Eldred2}
%V. D. Shiltsev, J. S. Eldred, V. A. Lebedev, and K. Seiya,
%\textquotedblleft{Beam Losses and emittance growth studies at the Record high space-charge dQ\_SC in the Booster}\textquotedblright,
%presented at the 12th Int. Particle Accelerator Conf. (IPAC'21), Campinas, Brazil, May 2021, paper WEXB08. 


%16
%\bibitem{IBIC21} V.Shiltsev, “Space-Charge and Other Effects in Fermilab Booster and IOTA Rings’ Ionization Profile Monitors”, presented at IBIC2021, Pohang, Rep. of Korea, 2021, paper TUPP05. 
%\url{doi:10.18429/JACoW-IBIC2021-TUPP05}.


%18
%\bibitem{Bakker} C.~Bakker and E.~Segre, ``Stopping Power and Energy Loss for Ion Pair Production for 340-Mev Protons", \emph{Phys. Rev.}, vol. 81, p. 489, 1951.
%\url{doi:10.1103/PhysRev.81.489}

%19
%\bibitem{Vilsmeier} D.~Vilsmeier, M.~Sapinski, and R.~Singh, ``Space-charge distortion of transverse profiles measured by electron-based ionization profile monitors and correction methods", \emph{Phys.Rev.Accel.Beams}, vol. 22, p. 052801, 2019.
%\url{doi:10.1103/PhysRevAccelBeams.22.052801}
%
%20
%\bibitem{Dimopoulou} C.~Dimopoulou \textit{et al.}, ``Dissociative ionization of H2 by fast protons: three-body break-up and molecular-frame electron emission", \emph{J.Phys.B: At. Mol. Opt. Phys.}, vol. 38, p. 593, 2005.
%\url{doi:10.1088/0953-4075/38/5/010}

	\end{thebibliography}
	{%
	% "biblatex" is not used, go the "manual" way
	%\begin{thebibliography}{99}   % Use for  10-99  references

%
% only for "biblatex"
%
%\ifboolexpr{bool{jacowbiblatex}}%
%	{\printbibliography}%
%	{%
	% "biblatex" is not used, go the "manual" way
	
	%\begin{thebibliography}{99}   % Use for  10-99  references
%	\begin{thebibliography}{9} % Use for 1-9 references
%	
%	\bibitem{jacow-help}
%		JACoW,
%		\url{http://www.jacow.org}
%	
%	\bibitem{IEEE}
%		\textit{IEEE Editorial Style Manual},
%		IEEE Periodicals, Piscataway,
%		NJ, USA, Oct. 2014, pp. 34--52.
%
%	\bibitem{journal-abbreviations}
%	\url{https://woodward.library.ubc.ca/researchhelp/journal-abbreviations/}
%
%	\end{thebibliography}
} % end \ifboolexpr
%
% for use as JACoW template the inclusion of the ANNEX parts have been commented out
% to generate the complete documentation please remove the "%" of the next two commands
% 
%%%\newpage

%%%\include{annexes-A4}

\end{document}